\documentclass{moriond}

\def\be{\begin{equation}}
\def\ee{\end{equation}}
\def\bea{\begin{eqnarray}}
\def\eea{\end{eqnarray}}

\begin{document}
\vspace*{4cm}
\title{Study of blazar variability through their ﬂux distribution for CTAO long-term monitoring}

\author{ A. Mikhno, J. Biteau, J.-P. Lenain on behalf of the CTAO collaboration}

\address{Sorbonne Université, CNRS/IN2P3, Laboratoire de Physique Nucléaire et de Hautes Energies, LPNHE,\\
4 place Jussieu, 75005 Paris, France}

\maketitle\abstracts{
The CTAO Key Science Project on Active Galactic Nuclei (AGN) includes a long-term monitoring (LTM) program of blazars. The probability distribution function (PDF) of the flux extracted from unbiased blazar lightcurves can provide important insight into the physical processes taking place in relativistic jets and their connection to observed variability. The aim of this work is to define an optimal observational strategy for the CTAO LTM program using the flux distributions as a quantitative metric. To achieve this goal, we develop a simulation-based analysis framework that enables a systematic comparison of different observational strategies.
Light curves were simulated as would be observed with the future CTAO for a representative set of AGN sources, following four different observational strategies within a fixed time budget. For each simulated light curve, different PDF models were fitted and their goodness of fit was evaluated. Based on the best-fit parameters, Monte Carlo simulations were generated to assess the ability of the method to recover the underlying PDF model. Since the true model used to generate the simulated light curves is known, this procedure allows us to quantify the discriminatory power of the algorithm and to identify which observational strategy maximizes model discrimination. Additionally, the algorithm has the potential of discrimination against different models for the flux distributions, such as Gaussian and lognormal. Once validated on simulations, the method is applied to archival lightcurves from existing gamma-ray instruments to extract flux PDFs and study the physical processes driving blazar variability.
The analysis framework has been developed and tested on simulated CTAO-like data, showing good stability and robustness. A study to optimize the observing strategy for the CTAO LTM is presented.}

\section{Introduction}
Active Galactic Nuclei (AGN) are extremely bright regions of galaxies, composed of an accretion disk feeding a super massive black hole (SMBH). About 10\% of all AGN have a jet of relativistic particles.  The orientation of a jetted AGN system towards the observer corresponds to the particularly interesting class of AGN, that are called blazars. Blazars are of special importance because they are some of the brightest objects and hence can be observed at large distances ($z>1$)\cite{AGN_1}. This makes AGNs prime targets for one of the key science projects of the Cherenkov Telescope Array Observatory (CTAO)\cite{CTAO_agn}. Currently under construction, the CTAO will explore the gamma-ray sky at energies above 30 GeV, reaching up to several hundred TeV. 
To achieve this, the observatory will be equipped with up to 37 Small-Sized Telescopes in the Southern Hemisphere, as many as 23 Medium-Sized Telescopes distributed across both sites to cover its core energy range, and four Large-Sized Telescopes stationed in the Northern Hemisphere.

The emission from blazars is highly variable at essentially all wavelengths, showing variability on different timescales, from few minutes up to years\cite{AGN_2}. Studying variability, one can characterize the physical processes, as its timescales and constrain the location of the emission region. 
In this work we focus on the long term monitoring, the concept of which is to regularly observe a set of the AGNs independently of their flux state in order to have an unbiased access to their duty cycle and locate the gamma-ray emission region within the AGN.

The Long-Term Monitoring (LTM) program was designed as part of the broader AGN Key Science Project. 14 sources were selected, representing a diverse range of AGN sub-classes. The aim of this program is to understand how variability differs across different source classes and to identify the most effective observational strategy for each, in order to efficiently use the scientific potential of the observations.

\section{Selection of the observation strategy for the Long Term Monitoring program}

In this work, the flux distribution is used as the metric for the selection of the best observational strategy, as the blazar flux distributions hints at the nature of variability. For example, lognormal distributions could indicate multiplicative processes\cite{Giebels}. Variability can originate from propagating fluctuations of the accretion rate within the disk, producing lognormal statistics in the accretion rate. Alternatively, the lognormal distribution can be explained as a hint of jet-internal multiplicative processes\cite{Tavecchio}. Furthermore, the Gaussian flux distribution suggests additive emission processes. In this case, variability could arise from stochastic fluctuations in particle acceleration and cooling rates within the jet\cite{Sinha}.
Furthermore, the heavy-tailed behavior of the flux distribution could point to burst-like or reconnection-driven events in the jet\cite{Biteau}.

Taking the flux distribution as a metric, the best observational strategy is selected as the one which allows for the discrimination between different flux models. Four strategies were tested: observations of 10 minutes approximately three times per week, 30 minutes once per week, 60 minutes twice per month, and 120 minutes approximately once per month. As a baseline, the LTM program has a fixed time budget limited to 26 hours per source per year. 

To perform this selection, simulated light curves as would be observed with the CTAO are required. The \textbf{CtaAgnVar}\cite{ctaagnvar} pipeline, built on  \textbf{Gammapy}\cite{gammapy}, simulates and reconstructs time-dependent phenomena in gamma-ray astrophysical sources. Originally designed for AGN variability studies, it supports a variety of time-dependent emission models, including various blazar models. 
A log-normal time series was used as the temporal model, while the spectral shape was modeled as a log-parabola with an exponential cut-off. In the study the Prod5 v0.1 version of the Instrument Response Function is used. After generating the injected spectral and temporal models, the CTAO observations were simulated. Light curves were produced for all sources and cadences, accounting for visibility conditions, with 100 realizations per source. \textbf{CtaAgnVar} fits several spectral models to each time bin and selects the best-fitting one, allowing precise light curve reconstruction while avoiding overfitting.

\subsection{Fitting the flux distribution}

\begin{figure}
\begin{minipage}{0.50\linewidth}
\centerline{\includegraphics[width=0.8\linewidth]{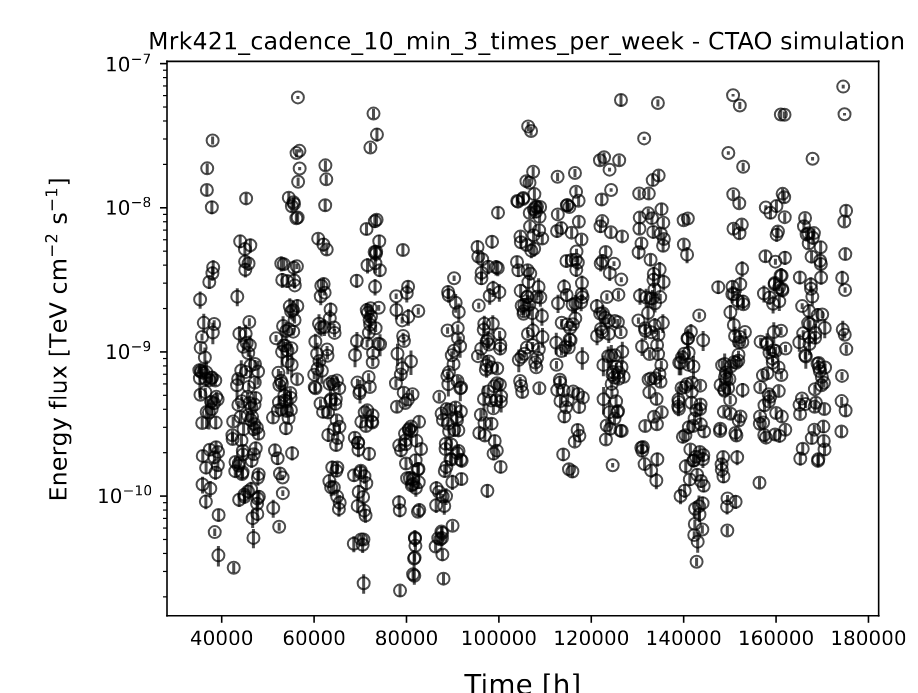}}
\end{minipage}
\hfill
\begin{minipage}{0.50\linewidth}
\centerline{\includegraphics[width=0.8\linewidth]{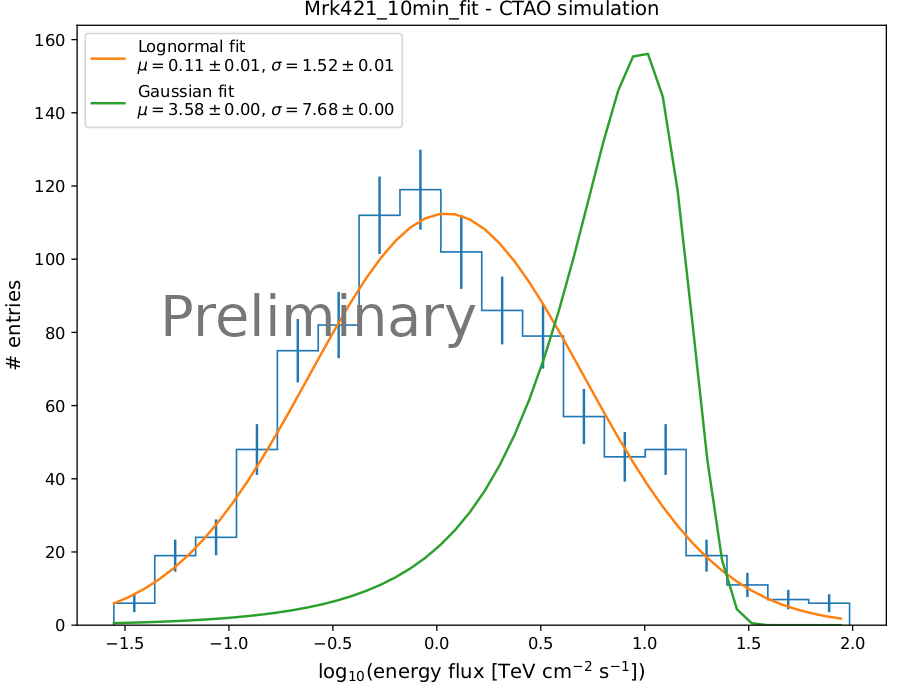}}
\end{minipage}
\hfill

\caption[]{Light curve for Markarian 421 with the cadence of observations of 10 minutes 3 times per week (left), and the fit of the corresponding flux distribution with Gaussian and lognormal models (right)}
\label{fig:LC}
\end{figure}

For each generated light curve, Gaussian and lognormal flux distribution models are fitted. Figure~\ref{fig:LC} shows an example of a light curve for Markarian 421 obtained from \textbf{CtaAgnVar} and the resulting fit. 

The resulting best-fit parameters are used to perform 1000 toy Monte Carlo (MC) simulations of light curves that follow the flux distribution.

\subsection{Sanity check and hypothesis test}
\begin{figure}
\begin{minipage}{0.45\linewidth}
\centerline{\includegraphics[width=0.95\linewidth]{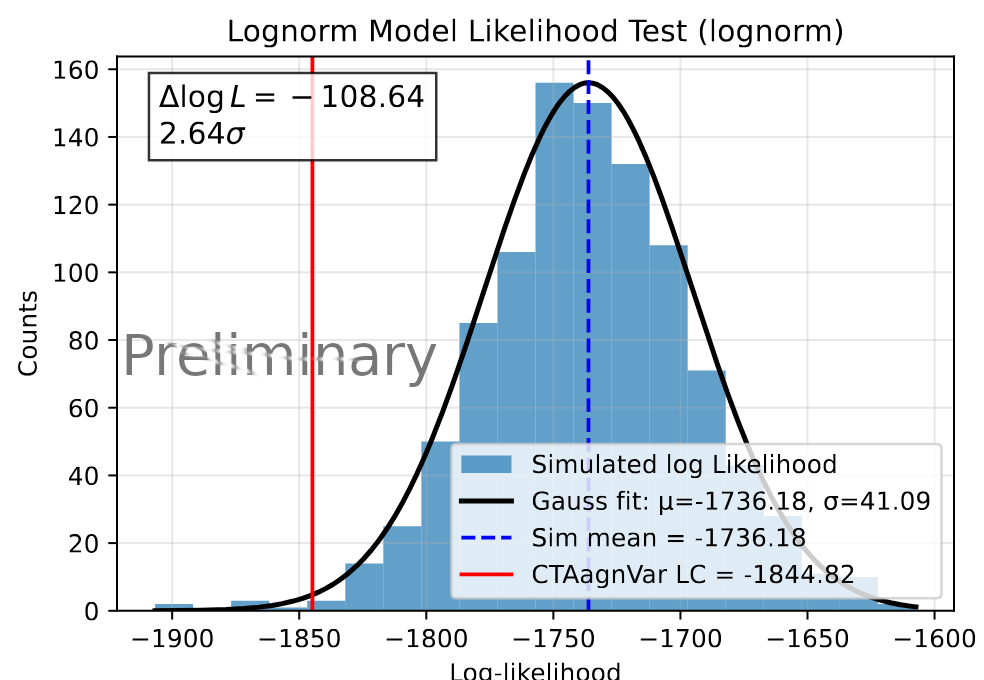}}
\end{minipage}
\hfill
\begin{minipage}{0.45\linewidth}
\centerline{\includegraphics[width=0.95\linewidth]{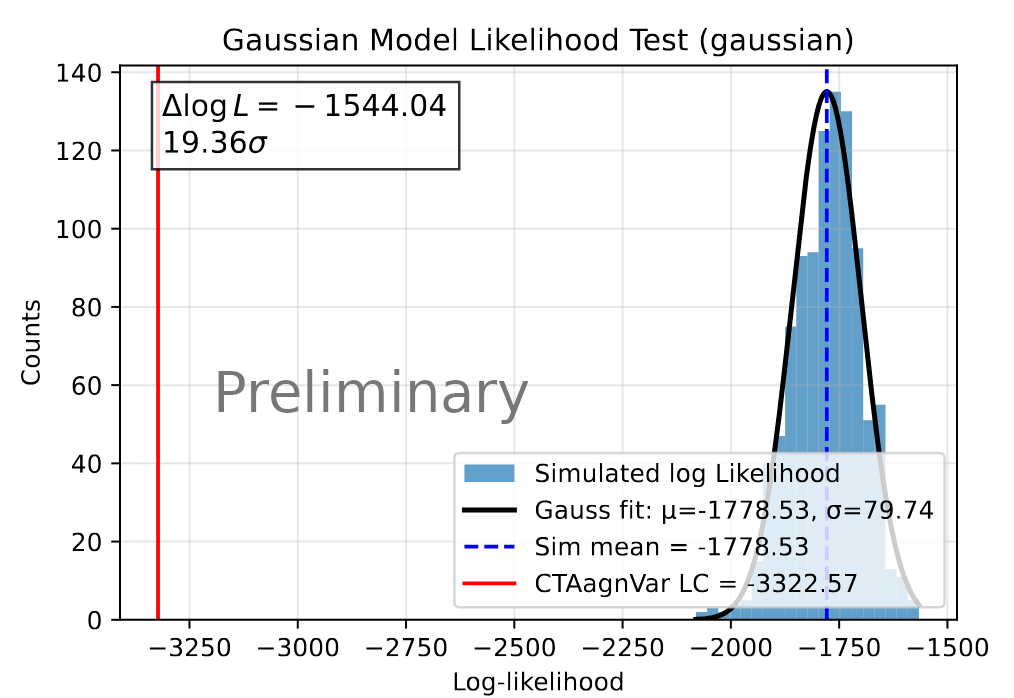}}
\end{minipage}
\hfill

\caption[]{Log Likelihood distributions for Markarian 421 with the cadence of observations of 10 minutes 3 times per week, for the initial light curves and toy MC simulations assuming lognormal flux distribution (left), and assuming gaussian flux distribution (right)}
\label{fig:test}
\end{figure}

By construction of the pipeline, the light curves generated using \textbf{CtaAgnVar} follow a lognormal flux distribution. Hence, the toy MC simulations generated using the best-fit parameters for the lognormal fit should reproduce the injected data. As a sanity check, the log-likelihood of the fit is computed for the initial light curve and for the toy MC. Then the significance with which an alternative hypothesis (Gaussian distribution) can be discriminated from the null hypothesis (log-normal distribution) is computed as:
\begin{equation}
    Z = \frac{\log \mathcal{L}_{\mathrm{obs}}   -\left\langle \log \mathcal{L}_{N,\mathrm{sim}} \right\rangle} {\sigma_{\mathrm{sim}}},
\label{eq:significance}
\end{equation}
where $\mathcal{L}_{\mathrm{obs}}$ is the likelihood for the initial simulated light curve,  $\left\langle \log \mathcal{L}_{N,\mathrm{sim}}\right\rangle$ is the mean over likelihoods for toy MC, and $\sigma_{\mathrm{sim}}$ is the standard deviation of the likelihood distribution of the MC simulations. Figure~\ref{fig:test} (left) illustrates the successful sanity check as the resulting significance value is $Z = 0.8\sigma$, which implies that the proposed algorithm allows to retrieve the initial flux distribution for the considered source at the considered cadence.

In order to test how accurately different underlying flux distributions can be discriminated depending on the cadence, the Gaussian distribution model was used for the toy MC simulations. The best-fit parameters from the fit to the \textbf{CtaAgnVar} light curve are used for the simulations. Knowing that the fluxes from \textbf{CtaAgnVar} follow a lognormal distribution, the significance level computed using equation~\ref{eq:significance} gives the discriminating power for the given cadence. For the considered case, the resulting significance is $Z=19\sigma$, which implies that it is possible to confidently distinguish between those two flux distribution models for this cadence.

\subsection{Selection of the best observational strategy}
\begin{figure}
    \centering
    \includegraphics[width=0.75\linewidth]{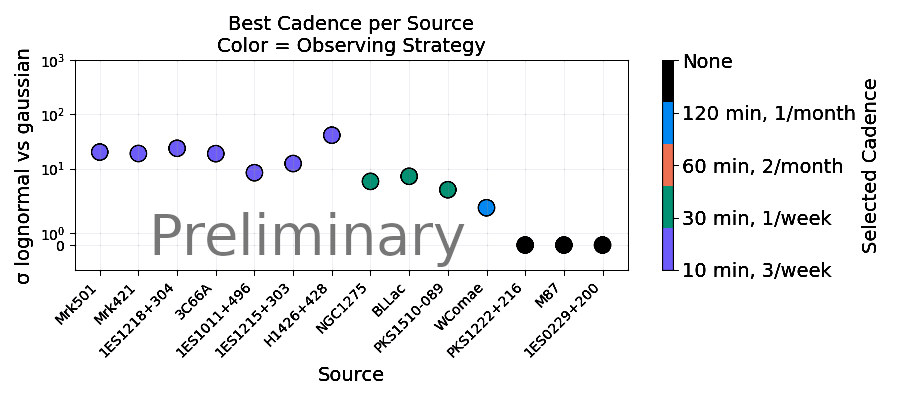}
    \caption{The observational strategies selected for each source}
    \label{fig:strategy}
\end{figure}

For each of the 14 sources of interest and 4 cadences, the hypothesis tests were performed following the procedure described above. A first condition requires the discriminating power to exceed $5\sigma$  for a cadence to be considered. However, this alone is insufficient, as for fainter sources observations of only 10 minutes may not provide significant detections of the source of interest. 
Hence, a second condition is introduced: at least 75\% of the considered time bins should correspond to at least a 3$\sigma$ detection of the source. This ensures that we are not only capable of discriminating between different flux models but also significantly detect the source in most of the observations. In Figure~\ref{fig:strategy}, the summary is presented with corresponding  discrimination significance between a Gaussian and a lognormal flux distribution and the selected cadences per each source. The best cadence depends on the brightness of the source, as for the brightest sources the optimal  strategy favors shortest observational windows, whereas for fainter sources longer windows are preferred. The faintest sources require particular attention, as none of the cadences yields a significant discrimination between flux distribution models.

\section*{Conclusion}
This work presents a methodology to devise the optimal observation strategy for the  CTAO LTM program. Using flux distributions as a discriminating metric, simulated light curves for 14 sources of interest were generated with \textbf{CtaAgnVar} and fitted with two flux distribution models. Testing alternative hypotheses on toy MC simulations, the discriminative power was quantified for each considered cadence, and the best observation strategy was selected for each source. 

\section*{Acknowledgments}
This work was conducted in the context of the CTA Consortium. We gratefully acknowledge financial support from the agencies and organizations listed here: $http://www.cta-observatory.org/consortium_acknowledgments$
\section*{References}
\bibliography{moriond}


\end{document}